\documentclass[amsmath,amssymb, aps,prl,twocolumn,notitlepage,nofootinbib]{revtex4-2} 
\usepackage{nameref}
\usepackage{graphicx}   
\usepackage{hyperref}
\usepackage{bm}
\usepackage{mathtools}
\graphicspath{{Figures/}}

\begin{document}
\title{Non-Fickian single-file pore transport}
\author{Spencer Farrell}
\author{Andrew D Rutenberg}
\email{adr@dal.ca}
\affiliation{
Dept. of Physics and Atmospheric Science, Dalhousie University, Halifax NS, Canada}%
\date{\today} 
\begin{abstract}
Single file diffusion (SFD) exhibits anomalously slow collective transport when particles are able to immobilize by binding and unbinding to the one-dimensional channel within which the particles diffuse. We have explored this system for short pore-like channels using a symmetric exclusion process (SEP) with fully stochastic dynamics. We find that for shorter channels, a non-Fickian regime emerges for slow binding kinetics. In this regime the average flux $\langle \Phi \rangle \sim 1/L^3$, where $L$ is the channel length in units of the particle size. We find that a two-state model describes this behavior well for sufficiently slow binding rates, where the binding rates determine the switching time between high-flux bursts of directed transport and low-flux leaky states. Each high-flux burst is Fickian with $\langle \Phi \rangle \sim 1/L$. Longer systems are more often in a low flux state, leading to the non-Fickian behavior. 
\end{abstract}
\maketitle

\textit{Introduction} --- Single-file diffusion (SFD), where particles diffuse within sufficiently narrow channels such that particles never exchange their relative positions, is exhibited within a variety of experimental systems \cite{Taloni2017} including zeolites \cite{Hahn1996}, colloidal racetracks \cite{Wei2000}, and carbon nanotubes \cite{Hummer2001}. SFD has also been well-studied theoretically in model one-dimensional systems, often focusing on the intriguing tracer dynamics of individual particles -- see e.g. \cite{vanBeijeren1983, Rodenbeck1999, Manzi2012, Leibovich2013, Ryabov2014, Krapivsky2014, Imamura2017}. Though in non-interacting SFD systems the collective diffusion of particles is the same as in simple diffusion (SD) \cite{Kutner1981, Farrell2015}, collective transport in SFD can be anomalously slowed with respect to SD due to transient immobility of particles due to binding and unbinding from channel walls \cite{Farrell2018}. 

Anomalous transport effects have been reported in single-file water transport through short channels. These effects range from bursty bidirectional transport seen in molecular-dynamics (MD) studies of short water-filled carbon nanotubes (CNT) \cite{Hummer2001} to experimental reports of an approximately exponential decrease of net flux with channel length in both biological and engineered  membrane pores \cite{Saparov2006, Horner2018}. While bursty transport in water-filled CNT has been attributed to the dynamics of chains of polarized hydrogen bonds spanning the length of the channel \cite{Waghe2002, Berezhkovskii2002}, the physics of the anomalous length dependence of transport in membrane pores is less clear. 

With simple diffusive systems, we expect Fick's law to hold, with average flux $\Phi$ inversely proportional to channel length $L$ -- i.e. $\Phi \sim 1/L$. This reflects a linear response of the flux to the average concentration gradient along the channel. Anomalous exponential scaling of flux with channel length \cite{Saparov2006, Horner2018} therefore represents ``non-Fickian'' behavior. Since these experimental studies are done with membrane pores with a geometric length spanning a phospholipid membrane, the exponential scaling is only uncovered when the effective length is measured by the number of charged residues within the channel. The anomalous length-dependence has therefore been  attributed to transient bonds between water and these charged residues \cite{Saparov2006, Horner2018}.

Since the reported non-Fickian length dependence in short charged channels requires interactions with the channel walls, we hypothesize that it may be related to anomalous transport in SFD due to binding and unbinding from channel walls \cite{Farrell2018}. Accordingly, we will stochastically model SFD transport in short one-dimensional channels. We will study the effects of transient particle immobility due to binding. 

We allow binding at every site of discrete one-dimensional channels, so our focus is on how channel-length $L$ affects the average flux of particles for fixed boundary concentrations -- i.e. how permeability scales with $L$. There are some limits in which we know that we will observe simple Fickian flux scaling $\Phi \sim 1/L$ \cite{Farrell2018}: for any length channel with either no particle association or with very fast association-dissociation dynamics then we expect SD behavior with Fickian transport, and for sufficiently long systems we also have anomalously slow but Fickian transport. The mechanism of this anomalously slow transport was found to be the trapping of mobile particles within ``cages'' of immobilized particles \cite{Farrell2018}. Since cages require multiple bound particles within a channel we expect different physics for smaller $L$. Our question then is, do short channels with slow binding-unbinding dynamics exhibit a distinctive transport regime? 

\begin{figure}[!t]
   \includegraphics[width=3.5in]{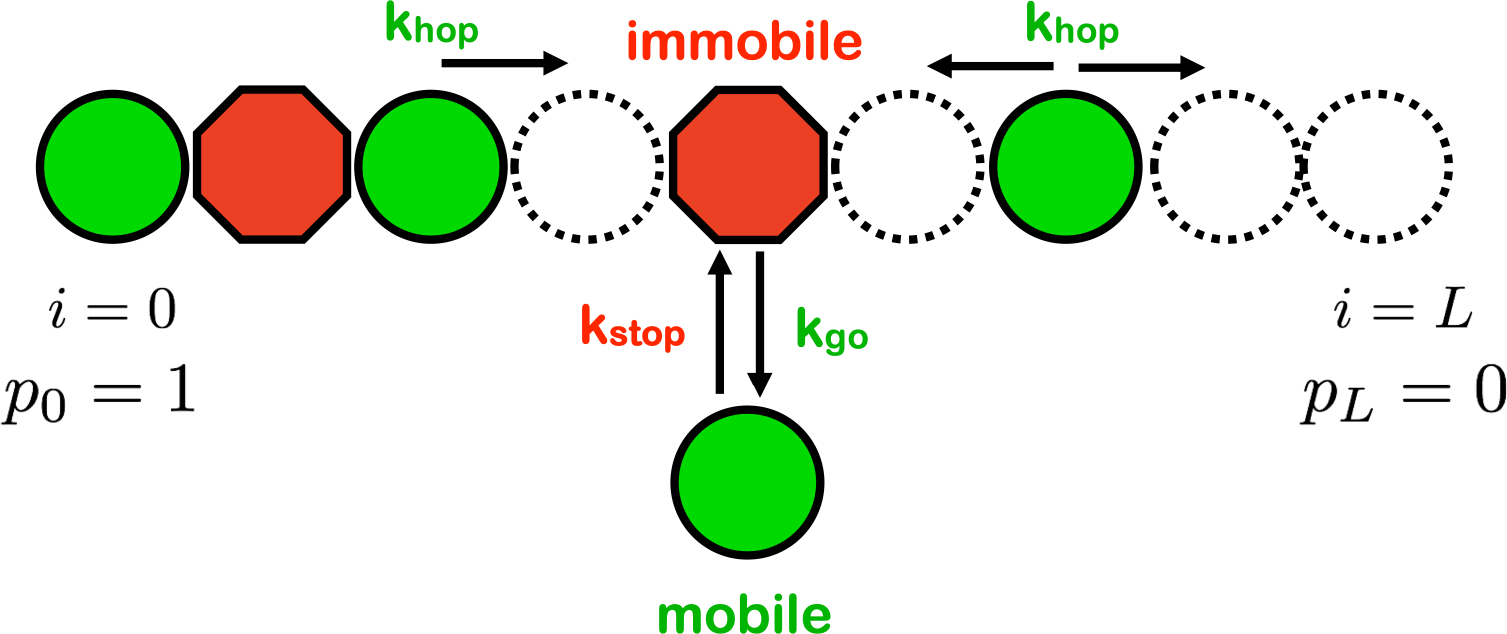}
    \caption{Schematic of our model system, with a one-dimensional channel of length $L$ indexed by discrete positions $i \in \{0,L\}$. Positions are either occupied (filled shapes) or unoccupied (dashed circles) by particles. A concentration gradient across the system induces an average net flux, and is maintained by imposing $p_0=100\%$ occupancy of the first site and $p_L=0\%$ occupancy of the last site, as indicated.  All particles can transition from mobile (green filled circles) to immobile (red filled octagons) states, or back, with transition rates $k_{stop}$ and $k_{go}$ respectively. The single-file constraint is imposed by allowing mobile particles to hop to an adjacent site at a rate $k_{hop}$ only if the destination site is not occupied.}  \label{fig:1} 
\end{figure}

\textit{Model} --- We consider a discrete symmetric exclusion process (SEP) within a one-dimensional channel with $L+1$ sites (we measure length in units of the site spacing $a$, which is an effective particle size), with occupancy probability $p$ that ranges from $p_0=1$ at the first site to $p_L=0$ at the last. As illustrated in Fig.~\ref{fig:1}, each particle can transition between mobile and immobile states at rates $k_{go}$ and $k_{stop}$, respectively. Mobile particles can hop into adjacent empty sites in either direction at rate $k_{hop}$. We simulate this system with an exact event-driven SSA (stochastic simulation algorithm, i.e. kinetic Monte Carlo) \cite{Gillespie1977}. We measure the average net flux across any point in the system in steady-state. Our code is freely available \cite{github}.

In this model isolated particles exhibit a diffusivity $D_0= k_{hop}/ \left[ 2 (1+K_A) \right]$, since they spend a fraction $1/(1+K_A)$ of their time bound, where $K_A=k_{stop}/k_{go}$ is the binding association constant (with disassociation constant $K_D =1/K_A$). For simple diffusion (SD) we would expect a flux $\Phi_0$ across the system of $\Phi_0 \equiv D_0 (p_0-p_L)/L$. We have previously shown that for $L \gg 1$ we observe a density dependent collective diffusion that is well described by $D=D_0/(1+\hat{p}+\hat{p}^2)$, where the occupation probability $p$ is scaled with $\hat{p} = p/p_{scale}$ and $p_{scale} \equiv \sqrt{(k_{stop}+k_{go})/k_{hop}} (1+K_D)$ is a characteristic density \cite{Farrell2018}.  This leads to an average flux
\begin{equation}
    \Phi_{Fick} = \frac{2 D_0 p_{scale}}{\sqrt{3} L} \left. \tan^{-1} \left[ (1+2 \hat{p}) \right] \right| ^{\hat{p}_0}_{\hat{p}_L},
    \label{eq:PhiLargeL}
\end{equation}
that exhibits Fickian ($\sim 1/L$) dependence on the system-size. In the limit of fast binding ($p_{scale} \rightarrow \infty$) $\Phi_{Fick} \rightarrow \Phi_0$, but for slower binding and unbinding kinetics the average flux is significantly smaller \cite{Farrell2018}. 

\begin{figure}[!t]
    \includegraphics[width=3.4in]{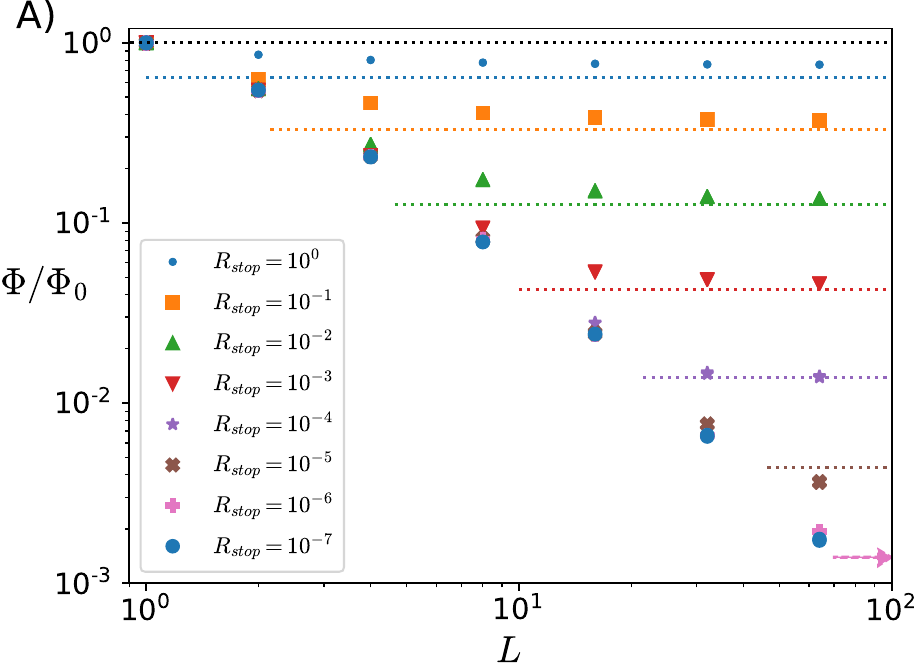}\\
    \includegraphics[width=3.4in]{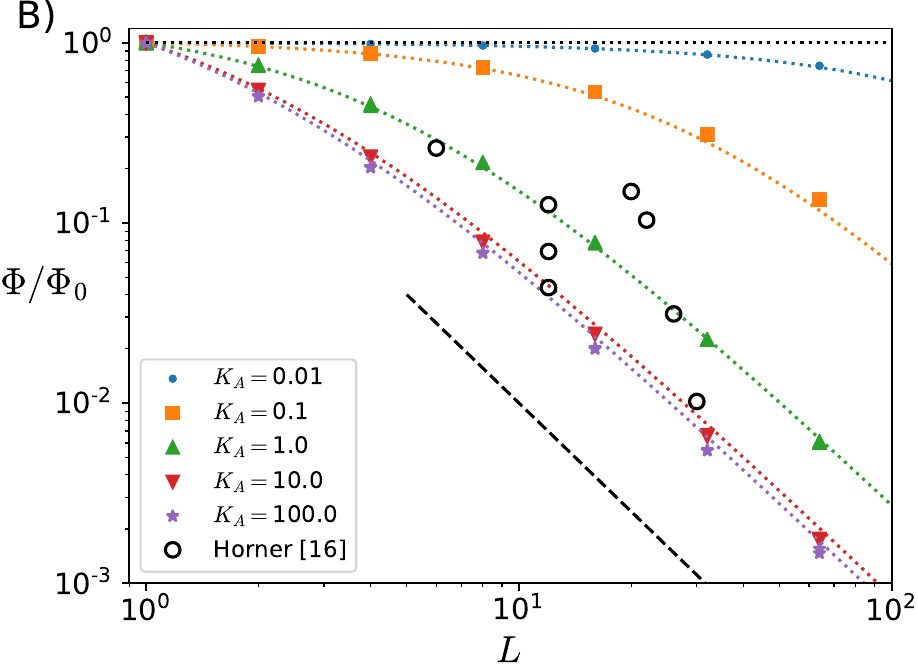} 
    \caption{The time-average flux $\Phi$ divided by the simple-diffusion (SD) flux $\Phi_0 = D_0/L$ vs. the system size $L$. A) For $K_A=10$ and various dimensionless binding rates $R\equiv k_{stop}/k_{hop}$ as indicated in the legend. The horizontal black dotted line is the SD result $\Phi/\Phi_0=1$, while the other horizontal dotted lines are the expected $L \gg 1$ (Fickian) behavior from Eq.~\ref{eq:PhiLargeL}, with colours corresponding to the legend. B) For $R_{stop}=10^{-7}$, for a variety of association constants $K_A$ as indicated. The diagonal dashed black line indicates $\Phi/\Phi_0 \sim 1/L^2$, i.e. $\Phi \sim 1/L^3$. The curved coloured lines are from the two-state model, Eq.~\ref{eq:phi}. The empty black circles are experimental data from \protect\cite{Horner2018}, see \emph{Discussion}.}    \label{fig:2}
\end{figure}

\textit{Results} --- In Fig.~\ref{fig:2}A, we plot the flux ratio $\Phi/\Phi_0$ vs the system size $L$ for a variety of $R_{stop} \equiv k_{stop}/k_{hop}$ ratios, as indicated by the legend. We use $K_A=10$, so that almost all particles are bound (i.e. immobile) at any given time. At $L=1$, solving the Master equation for this system exactly gives $\Phi/\Phi_0=1$ independently of either $R_{stop}$ or $K_A$ -- as observed. For fast binding dynamics, with $R_{stop} \gtrsim 1$, we also recover close to the SD flux ratio of $1$ \cite{Farrell2018}. For larger $L \gg 1$, we asymptotically approach the constant $\Phi_{Fick}/\Phi_0$ -- as indicated by the horizontal dashed lines \cite{Farrell2018}. Note that filled symbols represent time-averaged fluxes, ignoring the first half of data to minimize initial transients. Open symbols, behind the filled symbols, are averaged over the second quarter of the data -- their indistinguishability by eye indicates that steady state has been reached.

In Fig.~\ref{fig:2}A, we find that as $R_{stop}$ decreases the flux eventually becomes independent of $R_{stop}$.  As shown in Fig.~\ref{fig:2}B, we characterize this slow-binding regime for a wide range of $K_A$ values. Here, we use $R_{stop}=10^{-7}$ with coloured points as indicated. Since $\Phi_0 \sim 1/L$, the $L$ dependence observed in $\Phi/\Phi_0$ at intermediate channel lengths represents non-Fickian behavior. The heavy black dashed line indicates an approximate $1/L^2$ dependence of the ratio, implying $\Phi\sim 1/L^3$.  What is the mechanism of this behavior? 

\textit{Fast and slow flux states} --- We consider a two-state model, where our two flux states are fast (freely-flowing, denoted `f')  and slow (plugged, denoted `s'). The system transitions between these states due to the binding of particles within the channel. When particles are bound, they act as a plug that slows the flux. This model is described by the average lifespans of the states, $\tau_f$ and $\tau_s$, and their average fluxes, $\Phi_f$ and $\Phi_s$.

In the fast state, there are no bound particles in the channel and the average flux $\Phi_f = D_f (p_L-p_0)/L$, where $D_f \equiv k_{hop}/2$. The lifetime of this state before a binding occurs is $\tau_f = 1/( k_{stop} n_f ) \sim 1/L$, where the average number of particles is $n_f = (p_L+p_0)(L+1)/2$. This is a well defined dynamical state so long as $\tau_f$ exceeds the transient timescale needed to set up the constant density gradient, i.e. $\tau_f \gtrsim \tau_{trans} \equiv L^2/D_f$. This condition requires the channel to be shorter than a cross-over length
\begin{equation}
L_X \equiv R_{stop}^{-1/3}.
\end{equation}
Since $R_{stop}=k_{stop}/k_{hop}$, the $L \lesssim L_X$ regime extends to arbitrarily long pores in the limit of slow binding. For simplicity, here and subsequently we take $p_0=1$ and $p_L=0$ -- corresponding to our numerical boundary conditions.

In the slow state, there is at least one bound particle in the channel. We restrict our attention to the regime of the anomalous effects seen in Fig.~\ref{fig:2}B: $k_{hop} \gtrsim k_{stop} \gtrsim k_{go}$, i.e. slow but strong binding with $R_{stop} \lesssim 1$ and $K_A \gtrsim 1$. Initially we expect a plug-like configuration, with an average number of particles behind the first bound particle $n_{plug} = (L-1)/2$. With $K_A \gtrsim 1$, binding within the plug is faster than removing the first plugged particle and we expect $n_{bound} = 1+ \eta_s n_{plug}/(1+K_D)$ particles to be bound within the plug, where we introduce a dimensionless factor $\eta_s$. The lifetime of the plug is then $\tau_s \equiv n_{bound}/k_{go} \sim L$. Plugs are relatively long-lived, with $\tau_s > \tau_f$ when $K_A \gtrsim 1$. The slow flux state corresponds to slowly clearing the plug over $\tau_s$, so that $\Phi_s = (1+n_{plug})/\tau_s$.

With the lifetimes of the fast and slow flux states, $\tau_f$ and $\tau_s$, we also obtain their respective probabilities $q_f= 1/(1+\tau_s/\tau_f)$ and $q_s = 1-q_f$. The average flux in this two-state model is then $\Phi_2= q_f \Phi_f + q_s \Phi_s$. While the relative contribution of the slow flux increases with $L$, it is subdominant for $L \lesssim L_X$ where we obtain
\begin{eqnarray}
    \Phi_2 \simeq q_f\Phi_f &=& \frac{1}{1+\tau_s/\tau_f}\frac{k_{hop}}{2L}, \nonumber \\
    &=& \frac{k_{hop}}{2L\left[ 1 + 2n_{bound}/(k_{go}k_{stop}(L+1))\right]}, \nonumber \\
    &=&\frac{k_{hop}}{2L 
        \left[ 1 +  \frac{K_A (L+1)}{2} \left( 1+ \eta_s \frac{L-1}{2(1+K_D)} \right)  \right]}.
        \label{eq:phi}
\end{eqnarray}

For $L=1$ we recover the exact flux $\Phi=k_{hop}/\left[2(1+K_A) \right]$. The two-state model captures the observed fluxes for small $R_{stop}$ with $\eta_s = 0.55$, as shown by the dotted lines in Fig.~\ref{fig:2}B. For larger $L$, we observe $\Phi \simeq \Phi_2 \sim 1/L^3$ -- as indicated by the dashed black line. Note that $R_{stop}=10^{-7}$ in Fig.~\ref{fig:2}B, so that all $L \lesssim L_X$. When $L \gtrsim L_X$, indicated by the horizontal dashed lines in Fig.~\ref{fig:2}A, the flux crosses over to the Fickian behavior described by $\Phi_{Fick}$ in Eq.~\ref{eq:PhiLargeL}.

\begin{figure}[!t] 
    \includegraphics[width=3.5in]{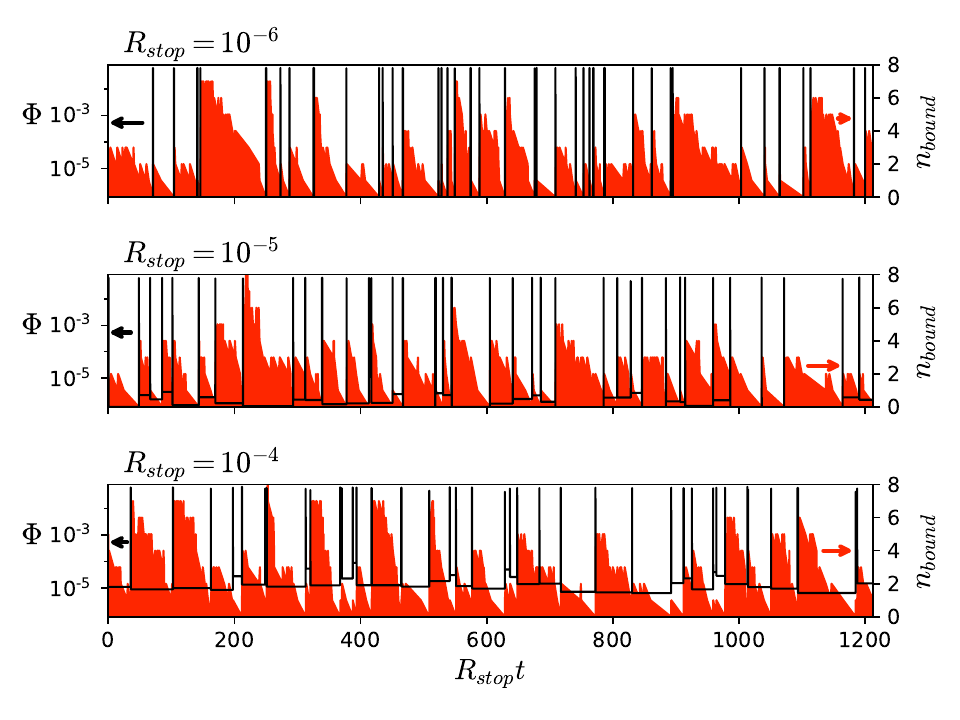}
    \caption{Two-state switching between fast and slow flux $\Phi$ (black traces, with a logarithmic scale on the left side) vs scaled time $t R_{stop}$. High flux coincides with no bound particles (red histograms, with a linear scale on the right side) while low flux coincides with bound (immobile) particles in the channel. Three different representative time-series are shown, with $R_{stop}=10^{-6}$, $10^{-5}$, and $10^{-4}$ (top to bottom, as indicated) at fixed $K_A = 10$ with $L=8$. The saw-tooth decrease of bound particle numbers between high-flux events indicates the slow clearing of the plug in the slow state, followed by a burst of flux in the fast state. Time units are such that $k_{hop}=1$.}
    \label{fig:3} 
\end{figure}

\begin{figure}[!t] 
\includegraphics[width=0.49\textwidth]{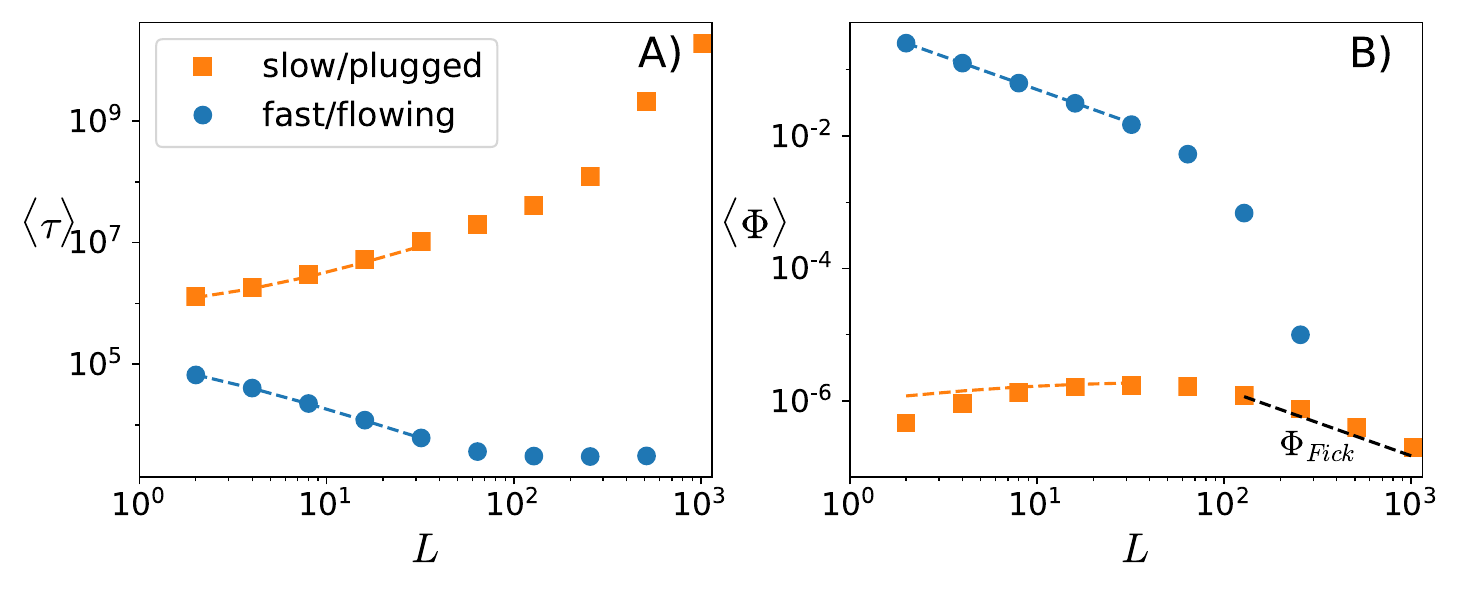} 
\includegraphics[width=0.49\textwidth]{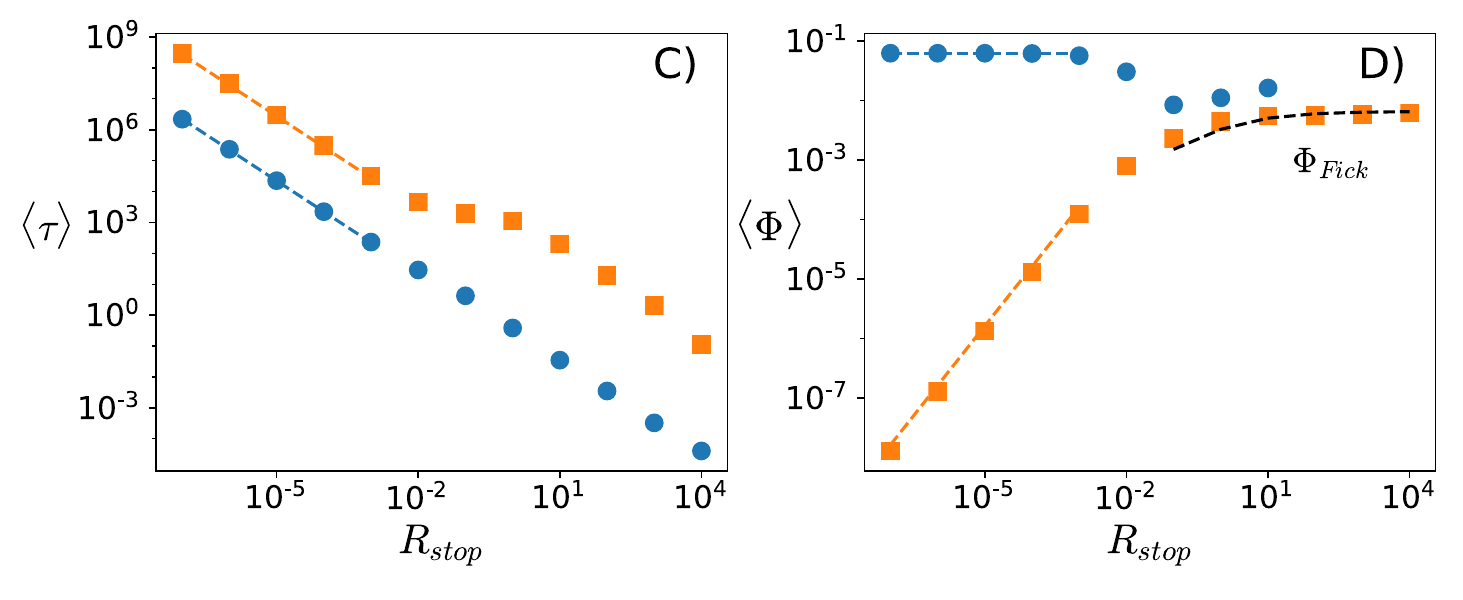} 
\caption{A) Average duration $\langle \tau \rangle$ of states with any particles bound (orange squares, slow) and all particles free (blue circles, fast) vs system length $L$. For all plots, $K_A=10$, $L=8$, and $R_{stop}=10^{-5}$, unless otherwise indicated; the two-state model calculations are indicated with coloured dashed lines for $L \leq L_X \equiv R_{stop}^{-1/3}$. B) Average flux $\langle \Phi \rangle$ in the slow (duration weighted) and fast states vs $L$. The dashed black lines indicates $\Phi_{Fick}$ from Eq.~\ref{eq:PhiLargeL}. C) Average duration $\langle \tau \rangle$ of slow and fast states vs $R_{stop}$.  D) Average flux $\langle \Phi \rangle$ for slow and fast states vs $R_{stop}$. Time units are such that $k_{hop}=1$.}    \label{fig:4} 
\end{figure}

\textit{Flux switching} ---  The speed of binding does not affect the average flux $\Phi_2$ in Eq.~\ref{eq:phi}, but determines the  largest channel length $L_X$ for which it is observed. Nevertheless, the lifetimes of the fast and slow states $\tau_f \sim 1/k_{stop}$ and $\tau_s \sim 1/k_{go}$ do depend on the binding rates. For fixed $K_A = k_{stop}/k_{go}$, this implies switching between high and low flux states with a rate proportional to the dimensionless binding rate $R_{stop} = k_{stop}/k_{hop}$ .

In Fig.~\ref{fig:3} we show time-series of flux (black lines, left axis) and number of bound particles (red histograms, right axis) vs scaled time $t R_{stop}$, for a hundred-fold variation of the dimensionless binding rate $R_{stop}$. The time-averaged flux values correspond to each distinct interval of time when the channel either has any or has no bound particles. These correspond to the slow and fast states, respectively, of our two-state model. 

The qualitative independence of the scaled plots in Fig.~\ref{fig:3} is consistent with the switching rate between states being scaled  by $R_{stop}$. The high-flux state also retains approximately the same magnitude as $R_{stop}$ is varied, in agreement with Eq.~\ref{eq:phi}. The system spends most of the time within the slow state, with a flux that remains orders of magnitude lower than the fast state but that increases with $R_{stop}$, and with a number of bound particles that decreases approximately linearly with time in a distinctive saw-tooth pattern. 
 
To examine this switching behavior more quantitatively, in Fig.~4 we plot both the average lifetimes $\langle \tau \rangle$ and the duration-weighted average fluxes $\langle \Phi \rangle$ of the two states (fast with no bound particles, or slow with bound particles) as either the system length $L$ or the dimensionless binding rate $R_{stop}$ is varied. We anticipate that our two-state model should apply for $L \lesssim L_X$, where we indicate the estimates of our two-state model with dashed lines. The two-state estimates of lifetime and flux are remarkably good.

For $L \lesssim L_X$, we observe $\langle \tau \rangle \sim 1/L$ for the free-flowing state, and $\langle \tau \rangle \sim L$ for the slow-bound state. The average flux $\langle \Phi \rangle$ when $L \lesssim L_X$ in the free-flowing state is Fickian with $\langle \Phi \rangle \sim 1/L$, indicating that the non-Fickian behavior of the system is controlled by the relative lifetimes of the free-flowing and slow-bound states. For smaller $R_{stop}$ we confirm that the average lifetime $\langle \tau \rangle \sim 1/R_{stop}$ for both states while the free-flowing flux is approximately independent of $R_{stop}$, in agreement with Eq.~\ref{eq:phi}.

For $L>L_X$, the average flux of the now-dominant slow-bound state is Fickian -- corresponding to $\Phi_{Fick}$ in Eq.~\ref{eq:PhiLargeL} \cite{Farrell2018} and indicated by the dashed black lines.  In this regime the fast state does not reach a dynamical steady-state, since $\langle \tau_f \rangle$ is independent of $L$ and $\langle \tau_f \rangle \lesssim \tau_{trans}$. The slow and fast fluxes approach each other both when $L \gtrsim L_X$, indicating that the dynamical transition out of the two-state flux-switching regime is continuous. 

\textit{Discussion} --- We have identified and characterized a collective transport regime for single-file diffusion (SFD) within short pore-like  channels. This regime is controlled by slow binding and unbinding of diffusing particles to the channel walls, i.e. transitions between mobile and immobile particles. The average particle flux is independent of the binding kinetics -- it only depends on the channel length $L$ and the association constant $K_A$ describing the equilibrium particle association. The binding kinetics determines the switching rate between two transport states: fast-flowing with no stationary particles, and slowly-leaking with at least one stationary particle jamming the pore. Transitions between the flux states are driven by binding of particles to the channel walls. While fast-flowing flux is Fickian (with $\langle \Phi \rangle \sim 1/L$) and dominates the average flux, the system is only in this state a fraction $q_f \sim 1/L^2$ of the time. As a result the average flux is non-Fickian with $ \langle \Phi \rangle \sim 1/L^3$ for $L \lesssim L_X$, where $L_X \equiv R_{stop}^{-1/3}$ is a crossover length determined by the dimensionless binding kinetics ($R_{stop} \equiv k_{stop}/k_{hop}$).

The anomalous pore-transport regime explored in this paper can be identified by three effects. First is the non-Fickian dependence on channel length $L$. Second is the distinctive random switching between bursts of high flux and a low flux state, with a time-scale that depends on both $L$ and $R_{stop}$. Third is the independence of the average flux on the binding rate $R_{stop}$. All of these effects are closely connected with the SFD nature of the system, and requires slow transitioning of individual particles between mobile and immobile states.

SFD effects in pore transport are important in the contexts of biological transport \cite{Bressloff2013}, nonequilibrium statistical mechanics \cite{Chou2011}, and nanofluidics \cite{Kavokine2021}. Many physical effects are not considered in our model, such as entry and exit dynamics \cite{Chou1998}, two-species or osmotic transport \cite{Chou1999}, collective or cluster diffusion \cite{Chou2011, Majumder2006}, or heterogeneities along the pore \cite{Kolomeisky2011}. These represent possible extensions of our model.

Molecular dynamics (MD) studies of SFD of water in CNT have reported non-Fickian scaling  of flow with length \cite{JingYuan2007}, but they have been limited to undriven systems without concentration or pressure gradients and no net flux. As a result, they reported \emph{absolute} rather than net flow -- as did  earlier MD reports of bursty flow in short fixed-length CNT \cite{Hummer2001, Waghe2002, Berezhkovskii2002}. These effects have been ascribed to collective reversals of the hydrogen bonding chain within the CNT. Furthermore, MD studies with a non-zero concentration gradient reported no non-Fickian effects \cite{Crozier2001, Zhu2002, Zhu2004, Portella2007}.

Nevertheless, non-Fickian dependence of collective transport has been reported in experimental studies of water in protein pores \cite{Saparov2006, Horner2018}.  An approximately 100-fold decrease of permeability $p_f$ is observed over a narrow range from $0-30$ in the number of binding sites along the pore \cite{Horner2018} (Fig 8). The open circles in Fig.~\ref{fig:2}B show that this data can be phenomenologically described by our model with $\langle \Phi \rangle \sim 1/L^3$. Here we have taken $L$ to be the number of hydrogen bonding sites along the pore, and plot the reported permeability $p_f$ multiplied by $0.0002 L$ (i.e. we pick $\Phi_0 \sim 1/L$ to fit).

Considering Fig.~2B, we also observe a similar decrease of relative flux if $K_A \approx 1$ and $R_{stop} \lesssim 10^{-4}$. For water, given a molecular diffusivity of $D \approx 5 \times 10^{-5} cm^2/s$ \cite{Horner2018} and molecular size $d \approx 3 \times 10^{-10} m$, we expect a hopping rate $k_{hop} \approx 10^{11}/s$. We would then require $k_{stop} \lesssim 10^7/s$, with $k_{go} < k_{stop}$. Reported hydrogen bond lifetimes for water associated with proteins are typically much less than $\approx 100ps$, which is far too short. While hydrogen bond lifetimes of trapped water within globular proteins can reach $10ms$ \cite{Carugo2015}, this is due to slow protein rather than slow water dynamics \cite{Persson2008, Persson2013}. Using our mechanism to explain the reported non-Fickian permeability of water pores would therefore require slow local pore conformational changes in response to local water occupancy. While none have been reported, MD studies of, e.g., 100ns \cite{Portella2007} may be too short to capture them. 

More generally, to clearly separate the timescales of local binding and transport probably requires collective dynamics within either the particles or the channel. In contrast, Brownian dynamics of particles in a fixed potential (see e.g. \cite{Lips2018}) would lead to similar effective unbinding and hopping rates -- i.e. $k_{go} \approx k_{hop}$ or $R_{stop}/K_A \approx 1$. In that limit we do not expect to see strong anomalous transport effects \cite{Farrell2018}. We require slow binding dynamics with respect to particle hopping.

For macromolecular SFD transport, long binding lifetimes with respect to diffusion are possible -- and our results could apply. For long channels with $L \gtrsim L_X$, Fickian transport is still expected though with anomalously reduced diffusion \cite{Farrell2018}. Pore-like structures in many cellular secretion systems (e.g. \cite{Costa2015}) could exhibit non-Fickian transport, or luminal diffusion within microtubules (e.g. \cite{Nihongaki2021}). In any case, non-Fickian SFD in short channels represents an interesting dynamical regime for transport within the field of non-equilibrium statistical mechanics \cite{Chou2011}.

We thank ACENET and Compute Canada for computational resources. ADR thanks the Natural Sciences and Engineering Research Council (NSERC) for operating Grant No. RGPIN-2019-05888. SF thanks NSERC for a CGSM fellowship.

\bibliography{refs}
\end{document}